\documentclass[useAMS,usenatbib]{mnras}
 \usepackage{graphicx}
 \usepackage{graphics}
 \usepackage{multicol}
 \usepackage{subfig}
 \usepackage{etex}
 \usepackage{setspace}
 \usepackage{float}
 \usepackage{pdflscape}
 
\def\apj{ApJ}%
\def\apjl{ApJ}%
\def\apjs{ApJS}%
\def\ao{Appl.~Opt.}%
\def\aap{A\&A}%
\def\mnras{MNRAS}%
\def\ssr{Space~Sci.~Rev.}%
\def\nat{Nature}%
\def\physrep{Phys.~Rep.}%
\title[]{Thermonuclear X-ray bursts in rapid succession in 4U 1636-536
with \emph{AstroSat}-LAXPC}
 \author[A. Beri et al.]
{Aru~Beri,$^{1,2,3}$
\thanks{a.beri@soton.ac.uk}
Biswajit~Paul,$^3$ 
J~S~Yadav,$^4$ 
H~M~Antia,$^4$  P~C~Agrawal,$^5$
\newauthor
R.~K~Manchanda,$^6$
Dhiraj Dedhia,$^5$ 
Jai~Verdhan~Chauhan,$^5$
Mayukh~Pahari,$^2$
 \newauthor
Ranjeev Misra,$^7$
Tilak~Katoch,$^4$ 
P. Madhwani,$^4$
Parag~Shah,$^4$
Varun,$^3$
Sujay Mate,$^{3,8}$
\\
 $^1$, DST-INSPIRE Faculty, Indian Institute of Science Education and Research (IISER) Mohali, 
Punjab 140306, India  \\
 $^2$, Physics \& Astronomy, University of Southampton, Southampton, Hampshire SO17 1BJ, UK \\
 $^3$, Raman Research Institute, Sadashivnagar, C. V. Raman Avenue, Bangalore-560 080, India.\\
 $^4$, Tata Institute of Fundamental Research, Homi Bhabha Road, Mumbai 400005, India.\\
 $^5$, UM-DAE Center of Excellence for Basic Sciences, University of Mumbai, Kalina, Mumbai 400098, India \\
 $^6$, University of Mumbai, Kalina, Mumbai 400098, India \\
 $^7$, Inter-University Center for Astronomy and Astrophysics, Ganeshkhind, Pune 411007, India \\
 $^8$, IRAP, Universit{\'e} de Toulouse, CNRS, UPS, CNES, Toulouse, France \\} 
\begin{document}
\pagerange{\pageref{firstpage}--\pageref{lastpage}} 
\maketitle
\label{firstpage}

\begin{abstract}
We present results from an observation of the Low Mass X-ray Binary 4U~1636--536 obtained with
the \emph{LAXPC} instrument aboard \emph{AstroSat}.
The observations
of 4U~1636--536 made during the performance verification
phase of \emph{AstroSat} showed seven thermonuclear X-ray bursts in a total exposure of
$\sim$~65~ks over a period of about two consecutive days.
Moreover, the light curve of 4U~1636--536 revealed the presence of a 
rare triplet of X-ray bursts, having a wait time of
about 5.5~minutes between second and the third bursts.
We also present results from time-resolved spectroscopy performed during these
seven X-ray bursts.~In addition, we have also
detected a transient Quasi-periodic oscillation~(QPO)
at $\sim$~5~Hz.~However, we did not find any evidence of kilo-hertz QPOs and/or X-ray burst
oscillations, perhaps
due to the hard spectral state of the source during this observation.

\end{abstract}

\begin{keywords}
accretion, stars: neutron, X-rays: binaries, X-rays: bursts, individual:~4U~1636--536
\end{keywords}
\section{Introduction}

Thermonuclear X-ray bursts (also known as Type-I X-ray bursts) are eruptions in X-rays that
can be observed from neutron star~(NS) low mass X-ray binaries~(LMXBs). These are
triggered in the NS envelope due to unstable thermonuclear burning of hydrogen and helium accreted from a low-mass stellar
companion \citep[e.g.,][]{Woosley76, Schatz98}.
During thermonuclear X-ray bursts, the X-ray luminosity increases many times
of the persistent level \citep[for reviews, see][]{Lewin93,Strohmayer06,Galloway08}. 
Within a few tens of seconds $10^{38}-10^{39}$ ergs of energy is released.
We do not discuss herein long bursts and superbursts that are comparatively
more energetic, $10^{40}-10^{42}$ ergs \citep[for details refer.,][]{Zand11}.

It is believed that most of the accreted fuel is burned during a burst.~Therefore,
the atmosphere must be completely replaced before a new burst can ignite \citep[see e.g.,][]{Woosley04,Fisker08}.
X-ray bursts with a recurrence time of hours to days have been 
detected from over 100 sources in our Galaxy \footnote{https://www.sron.nl/~jeanz/bursterlist.html} 
\citep[see e.g.,][]{Lewin93,Strohmayer06,Galloway08}.
However, bursts having short recurrence times varying between $\sim$~3 to 45~minutes have been observed from
15 sources \citep[see][for details]{Keek10}. \\

A large fraction of X-ray burst studies have been done with the 
\textsc{Proportional Counter Array}~(\textsc{PCA}) aboard \textsc{Rossi X-ray Timing Explorer}~(\emph{RXTE})
\citep[see e.g.,][]{Galloway08}.
The shortest recurrence time reported by \citet{Galloway08} in \emph{RXTE}-\textsc{PCA} light curves
is $\sim$~6.4~minutes.~\citet{Linares09} later, reported a short recurrence time of
5.4~minutes using \emph{RXTE} observations of 4U~1636--536.
\citet{Keek10} reported the shortest recurrence time between X-ray bursts 
from 4U~1705--44 to be $\sim$~3.8~minutes.
There also exist an exceptional source,~IGR~J17480--2446
that showed large number of recurring X-ray bursts during its outburst in 2010.
During this outburst, X-ray bursts with recurrence time as short as 3.3~minutes were also
observed \citep{Motta11,Linares12}. 
A study carried out by \citet{Boirin07} using a long X-ray observation of EXO~0748--676 with the \emph{XMM-Newton} observatory
discovered short recurrence time (8-20~minutes) bursts.
The same authors showed that short recurrence time bursts are less energetic compared
to the ``normal'' recurrence times (few hours).
They also found that short recurrence time bursts have
peak temperature lower compared to the ``normal'' recurrence times
bursts and the profiles
of such bursts lack 50~-~100~s long tail. 
An extensive study carried out by \citet{Galloway08} using \emph{RXTE}
data of 48 accreting neutron stars revealed that the short recurrence time bursts in many other sources also showed a behaviour
that is similar to that observed by \citet{Boirin07}. \\

Several possibilities have been discussed to explain short recurrence time
X-ray bursts. For instance, \citet{Fisker04}
 suggested a waiting point in the chain of nuclear reaction network. Very recently \citet{Keek17}
proposed an alternative model that show that 
there are left overs of the accreted material after the previous X-ray event
and these are transported by convection to the ignition depth, producing short recurrence time burst.  \\

The Large Area X-ray Proportional Counter (\textsc{LAXPC}) instrument on-board the Indian
multi-wavelength mission \emph{AstroSat} has timing capabilities similar to that of \textsc{PCA}
aboard \emph{RXTE} but has a larger effective area at higher energies (see next section for details).
\citet{Verdhan17} and \citet{Sudip18} reported detections of thermonuclear X-ray bursts
in the LMXB 4U~1728--34.~These authors also showed the presence of burst oscillations during the X-ray burst. \\

4U~1636-536, an extraordinary~(most luminous) atoll source \citep{Hasinger89} 
which exhibits a wide variety of X-ray bursts such as double-peaked 
X-ray bursts \citep{Sztajno85,Sudip06b}, triple-peaked X-ray bursts \citep{Van86,Zhang09}, 
superbursts \citep{Wijnands01,Kuulkers04}, short recurrence time bursts \citep{Keek10}
was observed with \textsc{LAXPC} as a part of the performance verification.
In this work, we report results obtained from the observations 
of 4U~1636--536 with the \textsc{LAXPC} instrument.~We found a total of seven thermonuclear X-ray bursts
in these observations of 4U~1636--536 and in particular, we observed the presence of short recurrence time bursts 
(a rare triplet). We also present the results obtained from the time-resolved spectroscopy 
performed during X-ray bursts. This is a technique which is often 
 employed for the measurement of neutron star parameters
 such as its radius \citep[see e.g.,][for details]{Lewin93,Galloway08,Sudip10}.

\section{Observations}	
\subsection{\emph{AstroSat}-\textsc{LAXPC}}

\emph{AstroSat} is the first Indian multi-wavelength astronomical satellite
launched on September 28, 2015. It has five instruments on-board
\citep{Agrawal06, Paul13}.~Large Area X-ray Proportional Counter~(LAXPC)
is one of the primary instruments aboard.
\textsc{LAXPC} consists of three identical proportional counters with a total
effective area of $\sim$6000~$cm^2$ at 15~keV.
The \textsc{LAXPC} detectors
have a collimator with a field of view of about $1^{\circ} \times 1^{\circ}$.
Each \textsc{LAXPC} detector independently records the time of arrival of each photon with a time resolution of 10~$\mu$s.
The energy resolution at 30~keV is about 15$\%$,~12$\%$,~11$\%$ for 
\textsc{LAXPC10},~\textsc{LAXPC20},~\textsc{LAXPC30} respectively \citep{Yadav16,Antia17}.

The three proportional counters of \textsc{LAXPC} are named as 
\textsc{LAXPC10}, \textsc{LAXPC20}, \textsc{LAXPC30}.
\textsc{LAXPC} detectors work in an energy range of 3-80~keV.
Each proportional counter unit has five layers, each with
12 detector cells \citep[for details see][]{Yadav16,Antia17}. \\

\subsection{Data}

\textsc{LAXPC} data was collected in two different modes:
Broad Band Counting~(BBC) and Event Analysis mode~(EA).
Data collected in the event mode of \textsc{LAXPC} 
contains information about the time, anodeID and Pulse Height~(PHA)
of each event.~Therefore, we have used
EA mode data for the timing and spectral analysis.
As a part of the performance verification, 
\emph{AstroSat} observed thermonuclear X-ray burst source 4U~1636--536
covering 22 orbits during 15-16 February 2016.~Observation details
are given in Table~\ref{obs}.

\begin{table}
\caption{\textsc{LAXPC} observations details of 4U~1636--536 starting from MJD~57433.25.
 In this \textsc{LAXPC} observation, other than the data gaps due to satellite passage through the 
 SAA region and occultation of the source by 
 the Earth, there was some additional data gap in orbit no 02092, around 90,000 seconds (Figure 2).}
\begin{tabular}{|c c c c|} 
 \hline
 Orbit~Number   & On-Source Exposure~(s)   \\ [0.45ex] 
 \hline\hline
 02078  & 1925    \\ 
 02079 &  2769    \\ 
 02080 &  2764   \\ 
 02081 &  2776    \\ 
 02082 &  2771   \\
 02083 &  2764    \\ 
 02084 &  2392     \\
 02085 &  2815     \\ 
 02086 &  3239    \\ 
 02087 &  3662   \\ 
 02088 &  3851   \\
 02089 &  3741    \\
 02090 &  3303   \\ 
 02091 &  2779  \\ 
 02092 &  736    \\
 02093 &  2357    \\ 
 02094 &  2394    \\
 02095 &  2737     \\
 02096 &  2739  \\ 
 02097 &  2733    \\
 02098 &  2436 \\ 
 02099 &  2862  \\ [0.9ex]

 \hline
\end{tabular}
\label{obs}
\end{table}


%
%
%

\section{Results}
\subsection{Light Curves}

\begin{figure*}
\begin{multicols}{2}
\includegraphics[height=3.5in,width=2.1\columnwidth]{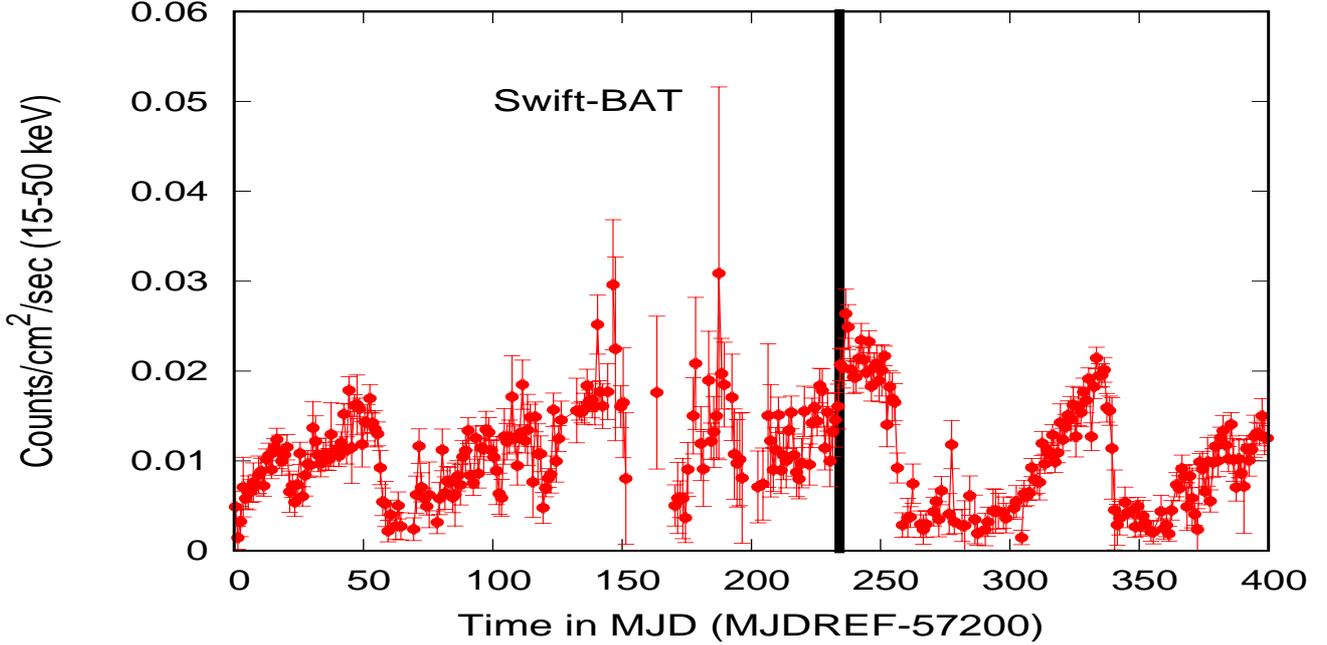}
\end{multicols}{2}
\caption{This plot highlights the spectral state of the source during the \textsc{LAXPC} observation using Swift-BAT light curve in 15-50~keV band.~Thick bold line 
indicates the time of the \textsc{LAXPC} observation~(around 57433~MJD).~This figure shows that \textsc{LAXPC} observation was made during the hard state of 4U~1636--536.}
\label{BAT}
\end{figure*}

\begin{figure*}
\begin{multicols}{2}
\includegraphics[height=2\columnwidth,width=1.5\columnwidth,angle=270]{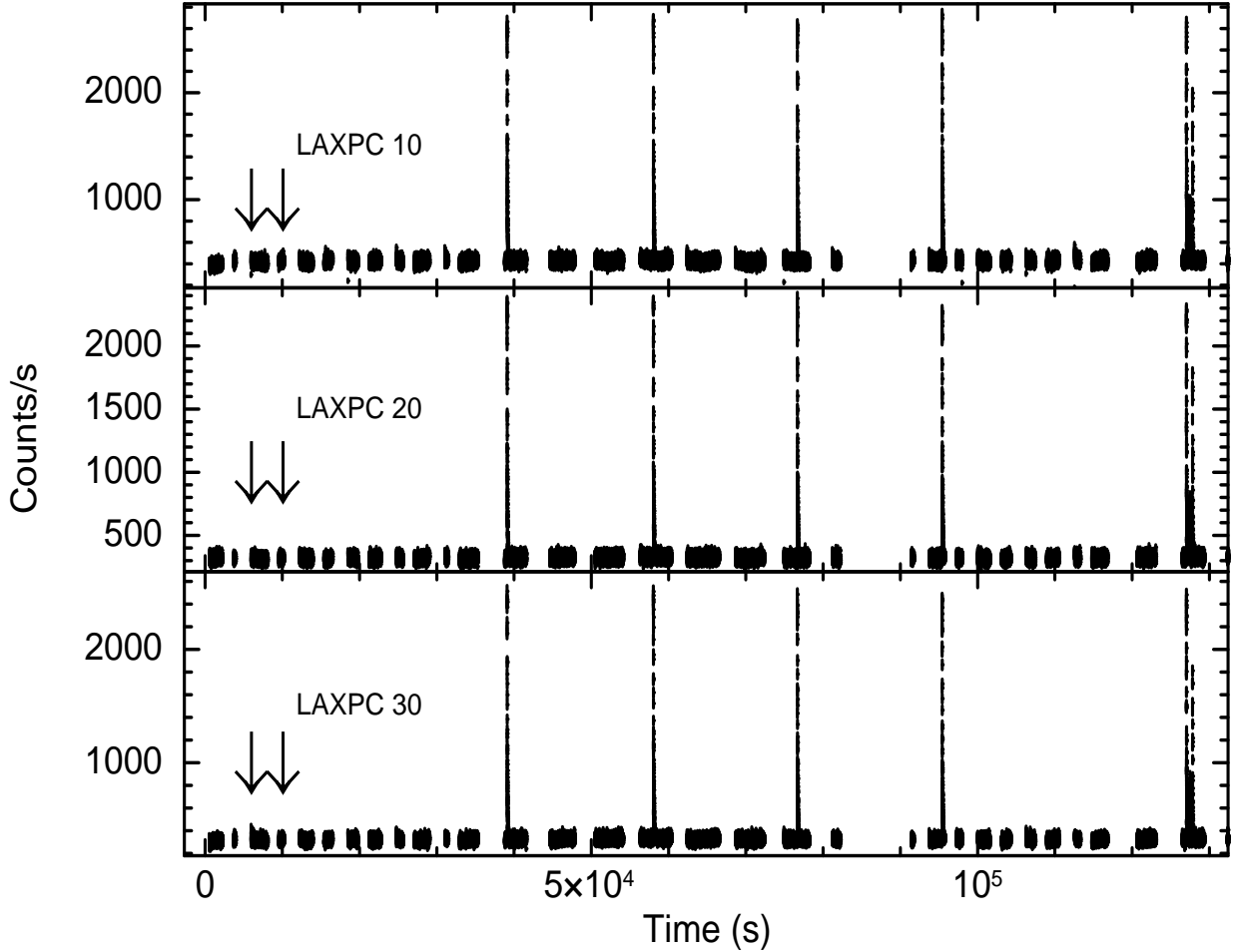}
\end{multicols}{2}
\caption{3-80~keV light curve created using \textsc{LAXPC} data of 4U~1636--536.
Light curve is binned with a binsize of 1~second.~The arrows indicate the segments of the light curve that show a low frequency QPO.}
\label{LC}
\end{figure*}

\begin{figure}
\includegraphics[height=\columnwidth,angle=270]{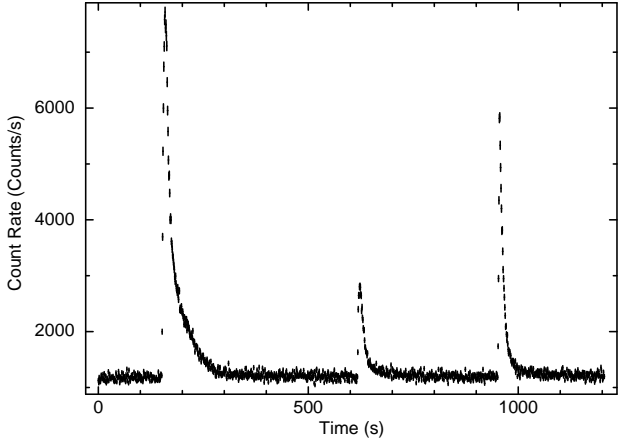}
\caption{3-80~keV light curve created using data of from the three proportional counters of \textsc{LAXPC}.
This plot shows the triplet of X-ray bursts (burst~5, burst~6 and burst~7).
Light curve is binned with a binsize of 1~second.}
\label{Triple}
\end{figure}

The \emph{Swift}/Burst Alert Telescope~(\textsc{BAT})
 is a hard X-ray transient monitor which observes
 $\sim$~88$\%$ of the sky each day and provides 
 near real-time coverage of the X-ray sky in the energy range
15-50~keV \citep{Krimm13}.~We have used 
\emph{Swift}-\textsc{BAT} light curve of 4U~1636--536 
to determine the spectral
state of this source (see Figure~\ref{BAT}).
The \textsc{BAT} light curve of 4U~1636--536 has been binned with a binsize of 1~day.
The \textsc{BAT} light curve of 4U~1636--536 shows
the long-term variability of about 40~days
from 2015 March to 2016 November.
Since \textsc{BAT} is a hard X-ray monitor therefore
the peak of the light curve~(high value of flux in 15-50~keV band)
would suggest the hard spectral state of a source. \\

Figure~\ref{LC} shows 3-80~keV light curve of 4U~1636--536 obtained using the data of \textsc{LAXPC}.
 We plot light curve of each detector separately to demonstrate the identical nature of light curves
in the three detectors of \textsc{LAXPC}.

For creating the light curve of 4U~1636-536 with \textsc{LAXPC},  
we have used data from all the layers of all the three proportional
counter units of \textsc{LAXPC}.
Good time intervals~(GTI) based on the following
filtering criteria:~elevation angle from Earth horizon greater than 2 degrees,
Charged Particle Monitor \citep[CPM;][]{Rao17} count rate less than 12~counts~s$^{-1}$ and angle offset less than 0.22 degrees
were used to create the source light curves.
We have used the \textsc{LAXPC} tool
task {\textquoteleft}{\textit{lxplc}}{\textquoteright} for the extraction of light curve.
\textit{lxplc} tool allows selection of events based on layers and energy channels.~We have
used channels ranging from 25-611 of each layer for the extraction of light curves of \textsc{LAXPC10},
channels 27-646 for \textsc{LAXPC30} while we have used channels 20-484 for the extraction of light curve using
data from \textsc{LAXPC20} (which has lower gain compared to \textsc{LAXPC10}, \textsc{LAXPC30} \citep{Antia17}.
The light curves were binned with a binsize of 1~second.
Light curve of each proportional counter unit consists of a persistent emission separated by data gaps and seven
thermonuclear X-ray bursts.
Data gaps in the observation corresponds to times
during the Earth occultation and the satellite passing through the South Atlantic Anomaly~(SAA) region.
The light curves show nearly a constant total count-rate of about 414 counts/s, 320 counts/s, 327 counts/s 
over the entire duration in \textsc{LAXPC10}, \textsc{LAXPC20}, \textsc{LAXPC30} respectively except during the 
seven thermonuclear X-ray bursts observed.
The average count-rate in \textsc{LAXPC10},
\textsc{LAXPC20}, \textsc{LAXPC30} measured during the earth occultation is 221 counts/s, 147 counts/s, 
137 counts/s respectively, which accounts for most of the background.
 
The X-ray bursts observed were found
in the data of orbits, viz. 02084, 02087, 02091, 02094 and 02099.
The zoomed-in version of these X-ray bursts in each  
\textsc{LAXPC} unit is shown in the Appendix~(Appendix A).
\subsection{A triplet of X-ray Bursts}

We have found
a very rare triplet of X-ray bursts in the light curve (see Figure~\ref{Triple}).
This triplet was seen in the last orbit of the observation.
To estimate the time difference or the wait time
between two X-ray events in a triplet, we followed
the definition of \citet{Boirin07}.~These authors
defined wait time
of a burst as the separation between its peak time and the
peak time of the previous burst.~The wait time between two X-ray bursts namely, burst 1 and
burst 2 of the triplet is $\sim$~7 minutes while the wait time between the last and the second burst
(burst 3 and burst 2 of the triplet) is about 5.5~minutes.

\subsection{Energy-Resolved Burst Profiles}
To observe the energy dependence
of these X-ray bursts, we created light curves during an X-ray burst in five different
narrow energy bands namely, 3-6~keV,~6-12~keV,~12-18~keV,
18-24~keV and 24-30~keV.~The light curves were created using data from all layers
of three counters of \textsc{LAXPC} and with a binsize of 1~second.
Figure~\ref{Burst-En} shows an energy resolved light curve of the
burst~1 seen in the observation of 4U~1636--536.
For the light curve in the 24-30~keV band, the count-rates have been multiplied
by a factor 5 for the visual clarity.~This figures indicates that 
bursts are detected upto 30~keV.~We extracted light curves 
above 30~keV to investigate the effect of bursts on the light curves at higher energies.
Light curves created in the 30-80~keV band showed the presence of a dip. The dip observed 
in the light curves was similar to the some reports of dip observed in hard X-rays during bursts
in other sources
\citep{Maccarone03,Chen13,Kajava17}.
However, dip observed in the \emph{LAXPC} light curves could
also be due to the deadtime effect and to investigate this we
corrected hard X-ray light curve for the total count rate in each detector.
The count rates were scaled by applying appropriate
deadtime correction{\footnote{http://www.rri.res.in/~rripoc/}}.
We observed after applying this correction the dip disappeared (see~Figure~\ref{DTC}).


We also noticed the energy dependence of the burst duration in Figure~\ref{Burst-En}.
Therefore, we measured the energy dependence of the burst duration
in all the seven X-ray bursts.
To measure the exponential decay times during each energy resolved
burst profile we fitted the burst profiles with the QDP model '\textit{burs}'.
We found that there is a gradual decrease
in the decay time with an increase in energy.
Decay time of an X-ray burst varied between $\sim$~28~seconds 
and 10~seconds (Figure~\ref{Decay}).
A gradual decrease in the temperature due to cooling of burning ashes
along the burst decay is the cause of observed
energy dependence of burst duration \citep[see e.g.,][]{Degenaar16}.

\subsection{Time-Resolved Spectroscopy}

\begin{figure}
\centering
\includegraphics[height=\columnwidth,angle=270]{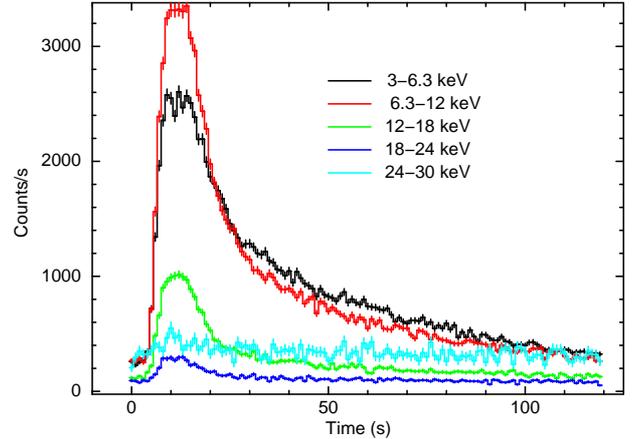}
\caption{This plot shows a thermonuclear X-ray burst as observed in different energy bands.~We have used data
during burst~1.~This figure shows that the X-ray burst is detected upto 30~keV.}
\label{Burst-En}
\end{figure}

\begin{figure}
\centering
\includegraphics[width=\columnwidth]{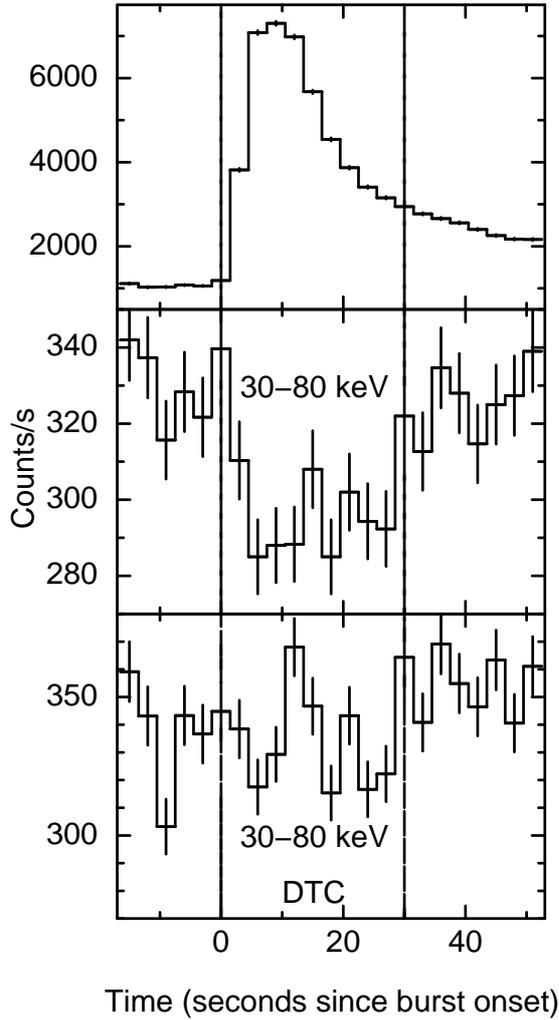}
\caption{This plot shows that the dip observed in the energy band of 30-80~keV close to the peak of thermonuclear X-ray burst~(burst~1)
disappears after performing deadtime correction.~The light curves are binned with a binsize of 3~s.}
\label{DTC}
\end{figure}

\begin{figure}
\centering
\includegraphics[width=\columnwidth]{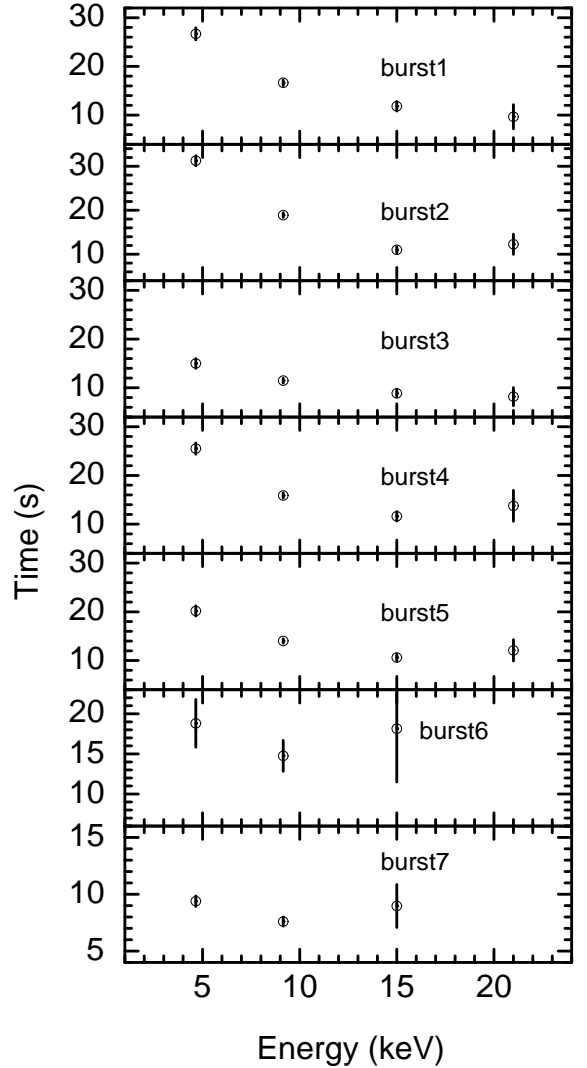}
\caption{This plot shows an exponential decay timescales observed in different energy bands during seven bursts observed
in the \textsc{LAXPC} data of 4U~1636--536.}
\label{Decay}
\end{figure}

We observed from the energy-resolved light curves that X-ray bursts are detected only upto 30~keV~(see Figure~\ref{Burst-En}).
In \textsc{LAXPC}, the soft and medium energy X-rays do not reach the bottom layers of the detectors.
Therefore, we have performed time-resolved spectroscopy using 
single events from the top layer~(L1,~L2) of each detector of \textsc{LAXPC}.~This was done to 
minimize the background.
Each proportional counter has a different energy response file
based on various factors gain, quantum efficiency, energy resolution etc. 
The gain of the \textsc{LAXPC} detector may vary from observation
to observation, therefore we first estimated the gain of each detector of \textsc{LAXPC}
using the tool $\textquoteleft$\textit{k-events-spec}'.
 We found that during the entire observations the value of gain compared to ground calibration data was
$\sim$1.0,~$\sim$0.99,~$\sim$1.05 for \textsc{LAXPC10}, \textsc{LAXPC20}, \textsc{LAXPC30} respectively.
Based on all these necessary checks we selected an appropriate response file
needed while performing spectroscopy.~Additional care was taken about the deadtime correction.
We modified the value of $\textquoteleft$\textit{backscal}' keyword
of each time-resolved spectra.
The value was calculated using $1.0/(1.0-N{\times}t)$, where $N$ is the count-rate and $t$ is
measured deadtime~($\sim$~42.3 microseconds).  \\

In order to investigate spectral evolution during X-ray bursts,
the spectra during X-ray bursts were extracted
using a time interval of 1~second.
For each burst, we extracted the spectrum
using 16~seconds of data preceding the burst.
This was subtracted for all burst intervals as the underlying 
accretion emission and background. \\

\begin{figure}
\includegraphics[height=\columnwidth,angle=-90]{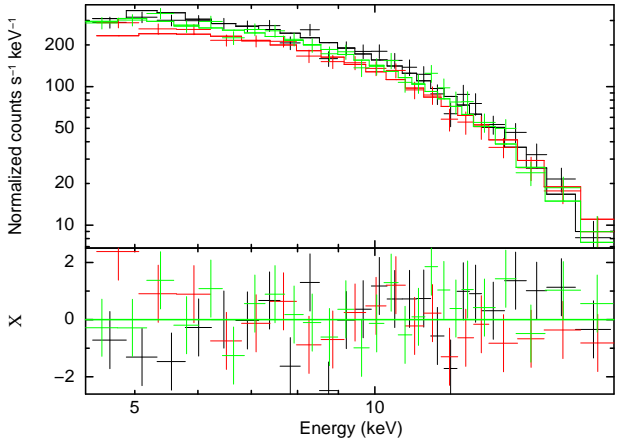}
\caption{An example best-fit count-rate spectrum of 4U~1636--536 obtained using the simultaneous spectral
fitting of the data of three detectors of \textsc{LAXPC}.~The lower panel shows
the residuals of the fit, where $\chi$ is defined as ($x_i$-$x_m$)/${\sigma}_i$.
$x_i$ and ${\sigma}_i$ corresponds to the observed counts and associated uncertainity while $x_m$
is the model prediction.}
\label{burst-spec}
\end{figure}

The net burst spectra in 4-20~keV band were fitted in XSPEC \citep{Arnaud96},
using a blackbody model {\textquoteleft}\textsc{BBODYRAD}' consisting
of two parameters, a temperature~($T_{BB}$) and a normalization
($K=(R_{BB}/d_{10})^2$), where $R_{BB}$ is the blackbody radius
and $d_{10}$ is the source distance in units of 10~kpc.~To account
for interstellar extinction, we used the '\textsc{TBABS}' model
component \citep{Wilms00}.~Since Hydrogen column density is not well constrained
in the \textsc{LAXPC} energy band, therefore we fixed $N_H=0.25{\times}10^{22}cm^{-2}$ \citep{Galloway08}.
Figure~\ref{burst-spec} shows an example count-rate
spectrum fitted with a blackbody model attenuated 
with an interstellar absorption component. 
While performing the spectral fitting we have also added a 
systematic error of $1{\%}$ to the time-resolved spectra. \\

%

\begin{figure*}
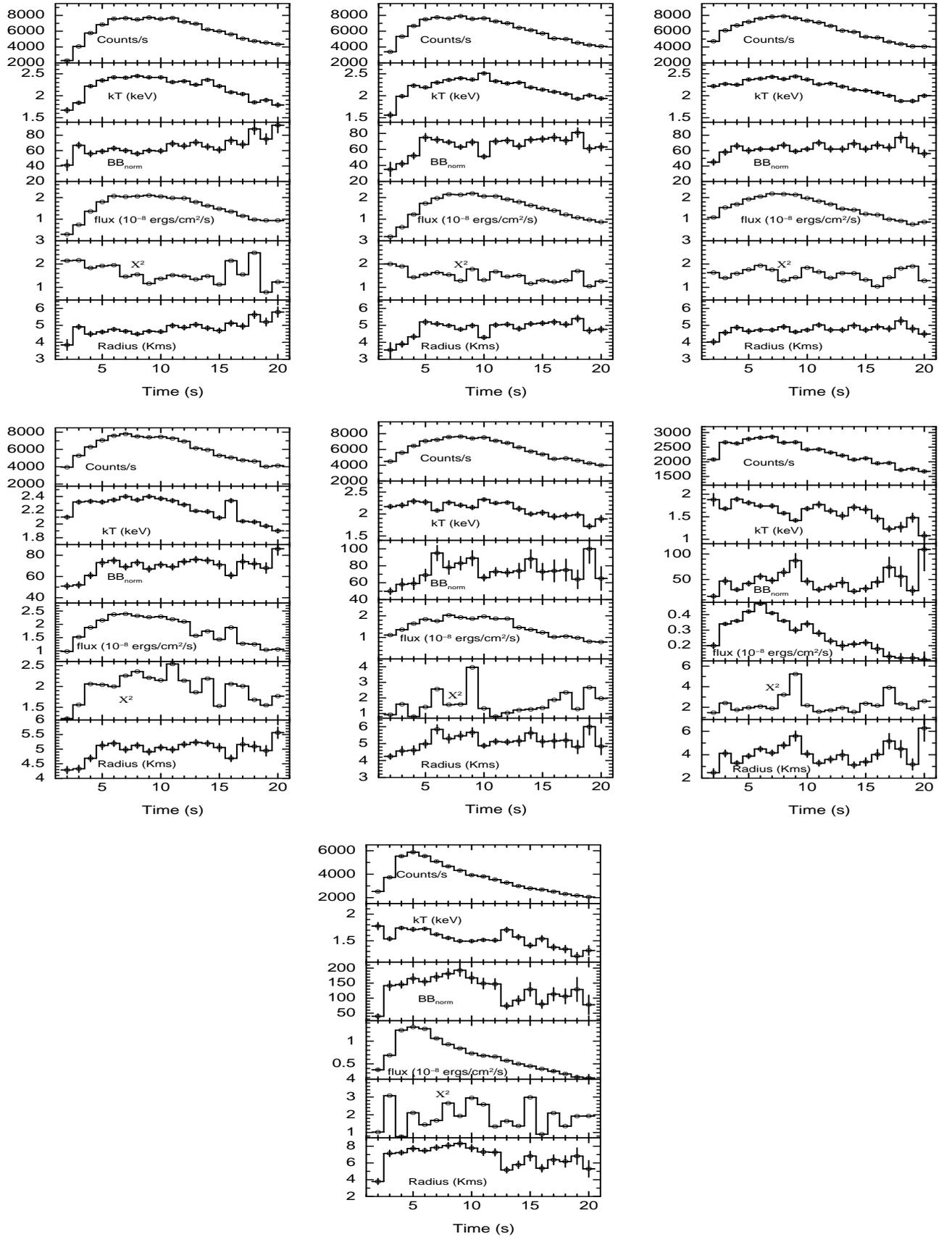

\centering
\begin{minipage}{0.3\textwidth}
\includegraphics[height=3.in,width=6.cm]{fig7.ps}
\includegraphics[height=3.in,width=6.cm]{fig10.ps}
\end{minipage}
\hspace{0.02\linewidth}
\begin{minipage}{0.3\textwidth}
 \includegraphics[height=3in,width=6.cm]{fig8.ps}
 \includegraphics[height=3in,width=6.cm]{fig11.ps}
\end{minipage}
\hspace{0.02\linewidth}
\begin{minipage}{0.3\textwidth}
 \includegraphics[height=3in,width=6.cm]{fig9.ps}
  \includegraphics[height=3in,width=6.cm]{fig12.ps}
 \end{minipage}
 \hspace{0.02\linewidth}
\begin{minipage}{0.3\textwidth}
 \includegraphics[height=3in,width=6.cm]{fig13.ps}
\end{minipage}
 \hspace{0.02\linewidth}
\caption{(First Panel)~In the left we show the best-fit parameters obtained after performing time-resolved spectroscopy of burst~1
.We have not applied correction factors to obtain a
real value of the neutron star radius.~Middle:~Best fit parameters obtained after performing time-resolved spectroscopy of burst~2
and the right plot shows the best-fit parameters obtained from the time-resolved spectroscopy of the burst~3.~(Second Panel)~In the left we show the best-fit parameters obtained after performing time-resolved spectroscopy \\
of the burst~4, the plot in the middle 
is for the burst~5 while the right plot is for burst~6.  \\
(Third Panel):~Best-fit parameters obtained after performing time-resolved spectroscopy of the burst~7.}
\label{Burst1}
\end{figure*}

In Figure~\ref{Burst1}, we show the best-fit
parameters obtained after performing time-resolved spectroscopy
of the seven bursts.~The top panel of each plot shows count-rate
during an X-ray burst, the second panel shows the temperature
evolution and the third panel shows the blackbody normalization
during each short segment of an X-ray burst.
Flux estimated in 4-20~keV band is shown in the fourth panel
of each plot while the fifth and the sixth panels show
the reduced chi-squared obtained from each spectral
fitting and the radius measured from the values of blackbody
normalization respectively.
We have not applied the correction factors such as color correction factor
for the measurement of a real value of neutron star radius.
The maximum temperature
 observed in an X-ray burst~1 is~$2.47\pm0.05$~keV.
 The emission radii inferred from the blackbody normalization
 spectral fit suggests that it is a Photospheric radius Expansion
 X-ray burst. Nearly same value of maximum temperature
 was also seen in burst~2,~3,~4 and~5.
However,~we notice a lower value of maximum temperature~($\sim$1.7~keV)
  in last two bursts of the triplet.

\begin{figure*}
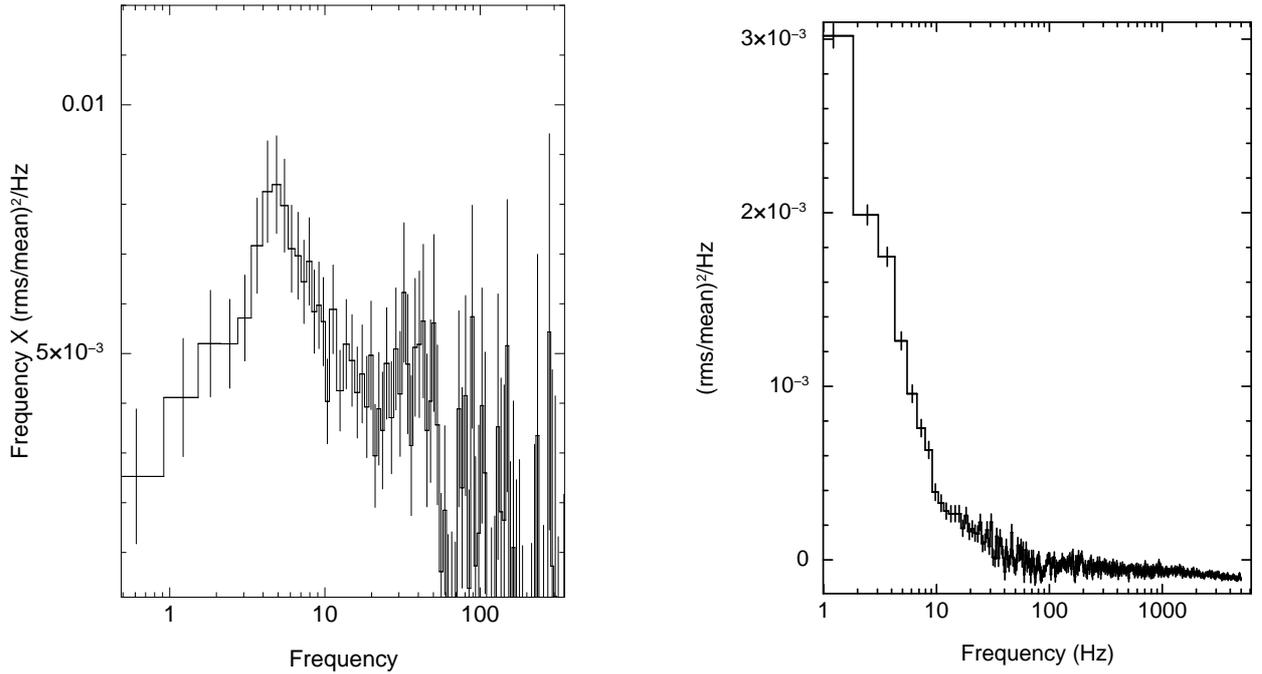

\centering
\begin{minipage}{0.45\textwidth}
\includegraphics[width=\columnwidth]{fig14.ps}
\end{minipage}
\hspace{0.05\linewidth}
\begin{minipage}{0.45\textwidth}
 \includegraphics[width=\columnwidth]{fig15.ps}
\end{minipage}
\caption{Left:~Power density spectrum created using combined light curve from LXP1, LXP2 and LXP3  of orbit -02079.
Light curves were created using data
from the top layers (L1,L2,L3 and L4) and in energy 2.5-25~keV.~There exist a weak QPO-like feature around $\sim$~5~Hz~(Q-value~$\sim$~1.2).~Right:~
We show the PDS covering wide frequency range 1-5000~Hz created
using the light curves of binsize 0.1ms. We have used data of all the detectors of \textsc{LAXPC}
and from all the orbits to obtain this PDS.}
\label{PDS}
\end{figure*}

\subsection{Power Density Spectrum}
We have added light curves from all the three detectors of \textsc{LAXPC} to 
generate the power density spectra of 4U~1636--536.
To have a high ratio of the source to the background
count rate,~we have used data from the top two layers and restricted
the light curves to 3-25~keV band\footnote{http://www.rri.res.in/$\sim$~rripoc/}.
We searched for both low frequency and high frequency
QPOs in the light curves excluding X-ray bursts.~For the case of low frequency QPOs we
used the light curves binned with a binsize 1~ms
while for the high frequency QPO search we used
light curves having binsize of 
0.1~ms.
The light curves were divided into segments each with 8192 bins.
PDS from all the segments were averaged to
produce the final PDS.
Figure~\ref{PDS} shows the presence of a weak low frequency QPO-like feature around 5~Hz
observed in the light curve of one of the orbit~(02079).
However, the Q-factor of this QPO-like feature is low $\sim$~1.2.
 We also searched for the presence of milli-hertz QPOs in the light curves of each orbit separately, however,
we did not find the presence of any such feature in the PDS.
On the right hand side of Figure~\ref{PDS}, we 
show the PDS obtained after using light curves from all the orbits.
This covers wide range of frequency 1-5000~Hz.
We also attempted to search for burst oscillations
using light curves during the X-ray bursts,
however, we did not find any burst oscillations
around the previously observed burst oscillation feature at 580~Hz~\citep{Strohmayer02}.

\begin{figure}
\includegraphics[width=\columnwidth]{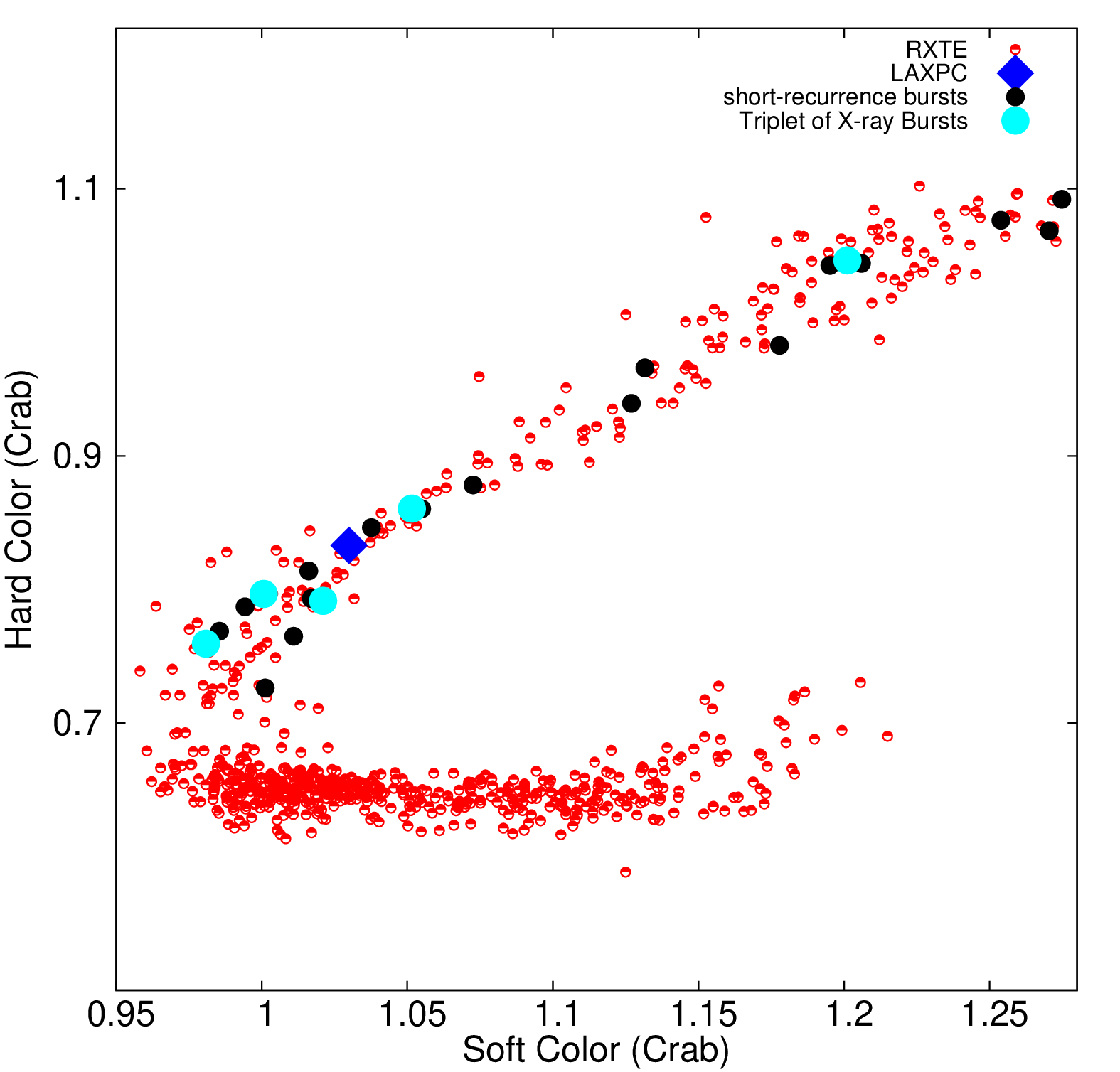}
\caption{Color-color diagram of 4U~1636--536 adapted from \citet{Altamirano08b}. 
(see text for the details about the energy bands used for the estimation
of soft and hard colors.).
This figure highlights the position
of \textsc{LAXPC} observation of 4U~1636--536 in the color-color diagram.
The observations performed with the \textsc{RXTE} observatory
that showed short recurrence X-ray bursts in this source are also shown.}
\label{color}
\end{figure}
\begin{table}
\caption{This table shows sources that exhibited series of three X-ray bursts 
having recurrence time less than 30 minutes.~We also note that \citet{Boirin07} 
reported five X-ray triplets in EXO~0748--676 having wait times of $\sim$12 minutes between the
three bursts of X-ray triplets.}
\begin{tabular}{|c c  |} 
 \hline
 
Source Name & Number of Triplets     \\ [0.45ex] 
 \hline\hline
 4U~1608--52$2^a$ &  1    \\ 
 EXO~1745--24$8^a$ & 1    \\
 2E~1742.9--292$9^a$ & 3 \\
 Rapid~Burste$r^a$ & 4 \\
 4U~1705--4$4^a$ & 1     \\
2S~1742--29$4^b$ & 3   \\
 Aql~X--$1^b$ & 1   \\
 EXO~0748--67$6^c$ &  5             \\[0.9ex]

 \hline
\end{tabular}
\\
{\bf{Notes}}: \\
   \hspace{0.7in} a $\rightarrow$ \citet{Galloway08} \\
     \hspace{0.7in} b $\rightarrow$ \citet{Keek10} \\
     \hspace{0.7in} c $\rightarrow$  \citet{Boirin07} \\

\label{Triplet}

\end{table}

%
\section{Discussions}
In this work, we have used a long \textsc{LAXPC} observation of 4U~1636--536.
The light curve of 4U~1636--536
revealed the presence of seven X-ray bursts.
Last orbit of the observation
showed a triplet of X-ray bursts.~Triplet of
X-ray bursts are rare events.
Therefore, we searched the literature and found that a total of 19
triplets of X-ray bursts have been observed so far from sources other than 4U~1636--536.
The wait time between the three bursts of these X-ray triplets 
is less then 30 minutes~(Table~\ref{Triplet}).~In particular, EXO~0748--676
showed X-ray bursts having
wait times of $\sim$12~minutes between the three bursts of triplets \citep{Boirin07}. \\

\begin{table}
\caption{Triplet of thermonuclear X-ray bursts observed in 4U~1636-536 with \emph{RXTE}~(\textsc{MINBAR} bursts).}
\begin{tabular}{| c c c |} 
 \hline
  Observation~ID & Wait Times & Wait Times \\ [0.45ex]
                 & (minutes)  & (minutes)  \\ [0.45ex]
                 & First \& second burst     &  Second \& third burst \\ [0.45ex]  
 \hline\hline
 
  60032-05-07-01 & $\sim$~8.4 & $\sim$~11.6 \\
  93087-01-42-10 &  $\sim$~13.8 & $\sim$~7.4  \\ [0.9ex]
  \hline
\end{tabular}

\label{Triplet-1636}

\end{table}
\subsection{Burst Catalog}

Multi-INstrument Burst ARchive (\textsc{MINBAR}) is a collection of 
Type~I bursts that are observed
with different X-ray observatories~(\emph{RXTE},~\emph{BeppoSAX},~\emph{INTEGRAL}), and that are all analyzed in a uniform way. 
Currently, it contains information on 6987 Type~I X-ray bursts from 84
burst sources (MINBAR version 0.9)\footnote{http://burst.sci.monash.edu/minbar}.
\textsc{MINBAR} was used to locate \emph{RXTE}-\textsc{PCA} observations of 4U~1636-536 
that shows bursts with recurrence times of less than 30~minutes.~Table~\ref{Triplet-1636} shows the X-ray burst
triplets observed in the \emph{RXTE}-\textsc{PCA} light curves and that were not included in \citet{Keek10}.
We found two additional triplet of X-ray bursts in 4U~1636-536 which were not
reported by \citet{Keek10}.
The shortest wait time observed between second and the third burst of a triplet of
X-ray bursts in the \emph{RXTE}-\textsc{PCA} light curves is 7.4~minutes.
However, the shortest wait time observed between
first and the second burst of a triplet in the \textsc{PCA} light curves is $\approx$~8~minutes. 
We thus, notice that the \textsc{LAXPC} observation of 4U~1636--536 revealed one of
the shortest triplet among all the known triplet of X-ray bursts~(Table~\ref{Triplet}) and also the 
shortest time difference ($\sim$ 5.5 minutes) among the triplet of bursts observed in this source (Table~\ref{Triplet-1636}).
Similar wait time was also reported by \citet{Linares09}
during the doublet of X-ray bursts of 4U~1636--536. 
Several models have been proposed to explain the occurence of short recurrence bursts.
Very recently, \citet{Keek17} proposed one-dimensional multi-zone model that reproduce short recurrence bursts
on the observed timescale of minutes.~Their simulations are based on the convective mixing driven by opacity.
The authors proposed that bursts that ignite in a relatively hot neutron star envelope 
leave a substantial fraction of unburnt fuel and convective mixing bring this fuel down to the ignition
depth on the observed timescale of minutes. 
We compared our observational results with that predicted by \citet{Keek17}.~We observe that although the recurrence 
timescales matches with that obtained by \citet{Keek17} but the profile of triple X-ray bursts obtained by these authors do not match
the one observed with \textsc{LAXPC}. The authors predicted intermediate bumps between the X-ray bursts which are caused by 
convection, however, we do not seen
such bumps in the light curves.~Moreover, the X-ray burst in the middle of the triplet is much fainter compared to the
other two X-ray bursts in the triplet. This is again not consistent with that predicted by \citet{Keek17}.
\citet{Grebenev17} recently studied short recurrence X-ray bursts using the data of \emph{INTEGRAL}. The authors
found the profile of triple X-ray bursts similar to that has been observed in 4U~1636-536 using the data of \textsc{LAXPC}.
The authors explained these short recurrence bursts based on the model of
a spreading layer of accreting matter over the neutron star surface. However, these authors believe that further refinement 
in the model is required for the complete understanding of the existence of triple X-ray bursts.

The large effective area of the \textsc{LAXPC} instrument at 
higher energies allowed us to create the energy resolved burst profiles.
We found that bursts were detected upto 30~keV.
Above 30 keV, the raw light curves of \textsc{LAXPC} detectors showed a dip similar to those 
reported in some other sources \citep{Maccarone03, Kajava17}. 
However, after a deadtime correction, 
the hard X-ray light curve during the burst was found to be consistent with the pre-burst emission.
We observe a strong energy dependence of burst decay times.
Decrease in the burst decay times with the increase in energy
suggests the cooling of burning ashes. \\

Time-resolved spectroscopy during X-ray bursts was performed
using spectra with 1~second time interval.~From the time
resolved spectroscopy we found that 
the maximum temperature measured is $\sim$ $2.5\pm0.05$~keV. \\

The power density spectral analysis showed the presence of a
QPO around 5~Hz.~The low frequency QPO is observed
in black holes and in accreting neutron stars as well
as in non-pulsating neutron stars. The low frequency QPO covers a wide range
between 0.1 and 50~Hz \citep[see][and references therein]{Altamirano08a}.
Moreover, these authors suggest that the range of low frequency covered
in a source is not related to spin
frequency, angular momentum or luminosity of the
object.
We did not detect any Kilo-hertz QPOs
in this observation of 4U~1636--536.
This again suggests hard spectral state of the source
as high frequency QPOs are often observed during the 
soft spectral state of this source \citep{Belloni07}.~This is well supported by the
fact that the \emph{Swift}-\textsc{BAT} light curve suggested
that this observation of 4U~1636--536 was 
made near the peak of its light curve in 15-50~keV band. \\

The Atoll sources are known to display three main tracks in the color-color diagram.
These three spectral states are: the extreme island state~(EIS),
the island state~(IS) and the banana branch~(BB) \citep{Hasinger89}.
\citet{Altamirano08c} have described in detail the characteristics of each 
of these states.~To confirm the hard~(island)~spectral state
of the present \textsc{LAXPC} observation, 
we have used the color-color diagram of 4U~1636--536 (Figure~1 of \citet{Altamirano08b}).
The authors defined soft and hard colors as
the ratio of count-rates in 3.5-6.0~keV / 2.0-3.5~keV
and 9.7-16.0~keV/6.0-9.7~keV respectively.
They also normalized the colors by the corresponding
Crab Nebula color values that are closest in time to
correct for the gain changes as well as for the differences in
the effective area between different proportional counters of \emph{RXTE}.
We have adapted the same technique to estimate the colors of the \textsc{LAXPC}
observation, however, we have used slightly different energy bands for the estimation of colors.
For the soft and the hard colors, we used the ratios of count-rates in 6.0-9.7~keV / 3.0-6.0~keV and 
16.0-25.0~keV / 9.7-16.0~keV respectively.
It is quite interesting to notice that the \textsc{LAXPC} observation falls on the island branch
which corresponds to the hard spectral state (see Figure~\ref{color}). 
We also noticed that all the previous known short recurrence bursts 
from this source \citep[see e.g.,][]{Keek10} lie on the island branch of the color-color diagram.
The observations performed with \emph{RXTE} that exhibit triplets of X-ray bursts~(shown in cyan color in Figure~\ref{color})
also lie close to this observation performed with \textsc{LAXPC}. \\

\citep{Altamirano08b} showed that close to the transition between the island 
and the banana state, 4U~1636--536 exhibits millihertz quasi-periodic oscillations~(mHz QPOs)
(see Figures~1 \& 2 of their paper). We did not detect any milli-hertz oscillations
in the power density spectrum created using the \textsc{LAXPC} observation.
The non-detection of mHz QPOs further supports the fact that the current
observation does not lie close to the transition region between the island 
and the banana state.
It is proposed that mHz QPOs are the consequence of marginally stable 
burning on the neutron star surface \citep{Heger07}.
The non-detection of mHz QPOs in the \textsc{LAXPC} observation 
suggests that the accreted fuel is not available for the
stable burning \citep{Altamirano08b}.~This is well supported by the fact that we observe
seven consecutive bursts in almost day long observation. \\

To summarize, we present the results obtained with the study of seven thermonuclear
X-ray bursts observed with \emph{AsroSat}-\textsc{LAXPC}.
The short recurrence time
bursts (a rare triplet) in the LMXB 4U~1636--536 have been detected for the first time with \emph{AsroSat}-\textsc{LAXPC}.
We have shown the energy dependence of burst profiles and have also discussed the caveats
due to the background at higher energies.
This paper along with two other papers on thermonuclear X-ray bursts observed with \textsc{LAXPC} \citep{Verdhan17,Sudip18}
demonstrates the sensitivity of the \textsc{LAXPC} instrument to detect and study thermonuclear X-ray bursts.

\section*{Acknowledgments}
A.B. \& M.P. gratefully acknowledge the Royal Society and SERB~(Science $\&$ Engineering Research Board, India)
for financial support through Newton-Bhabha Fund.
A.B thanks
Dr.~Diego Altamirano for providing the \textsc{RXTE} data for Figure~\ref{color}, and
for useful discussions.~We acknowledge strong support from Indian Space Research Organization (ISRO) in various aspect of  
\emph{LAXPC} instrument building, testing, software development and mission operation during payload verification phase.
This paper uses preliminary analysis results from the Multi-INstrument Burst ARchive (MINBAR), 
 which is supported under the Australian Academy of Science's Scientific Visits to Europe program, 
 and the Australian Research Council's Discovery Projects and Future Fellowship funding schemes.
 We also thank Dr.~Laurens Keek for providing information on bursts that were not included in their paper
 \citep{Keek10}.

%
%

\newpage

\bibliography{complete-manuscript}{}

\begin{thebibliography}{}
\makeatletter
\relax
\def\mn@urlcharsother{\let\do\@makeother \do\$\do\&\do\#\do\^\do\_\do\%\do\~}
\def\mn@doi{\begingroup\mn@urlcharsother \@ifnextchar [ {\mn@doi@}
  {\mn@doi@[]}}
\def\mn@doi@[#1]#2{\def\@tempa{#1}\ifx\@tempa\@empty \href
  {http://dx.doi.org/#2} {doi:#2}\else \href {http://dx.doi.org/#2} {#1}\fi
  \endgroup}
\def\mn@eprint#1#2{\mn@eprint@#1:#2::\@nil}
\def\mn@eprint@arXiv#1{\href {http://arxiv.org/abs/#1} {{\tt arXiv:#1}}}
\def\mn@eprint@dblp#1{\href {http://dblp.uni-trier.de/rec/bibtex/#1.xml}
  {dblp:#1}}
\def\mn@eprint@#1:#2:#3:#4\@nil{\def\@tempa {#1}\def\@tempb {#2}\def\@tempc
  {#3}\ifx \@tempc \@empty \let \@tempc \@tempb \let \@tempb \@tempa \fi \ifx
  \@tempb \@empty \def\@tempb {arXiv}\fi \@ifundefined
  {mn@eprint@\@tempb}{\@tempb:\@tempc}{\expandafter \expandafter \csname
  mn@eprint@\@tempb\endcsname \expandafter{\@tempc}}}

\bibitem[\protect\citeauthoryear{{Agrawal}}{{Agrawal}}{2006}]{Agrawal06}
{Agrawal} P.~C.,  2006, \mn@doi [Advances in Space Research]
  {10.1016/j.asr.2006.03.038}, \href
  {http://adsabs.harvard.edu/abs/2006AdSpR..38.2989A} {38, 2989}

\bibitem[\protect\citeauthoryear{{Altamirano}}{{Altamirano}}{2008}]{Altamirano08c}
{Altamirano} D.,  2008, PhD thesis, Sterrenkundig Instituut ``Anton Pannekoek''
  University of Amsterdam

\bibitem[\protect\citeauthoryear{{Altamirano}, {van der Klis}, {Wijnands}  \&
  {Cumming}}{{Altamirano} et~al.}{2008a}]{Altamirano08b}
{Altamirano} D.,  {van der Klis} M.,  {Wijnands} R.,   {Cumming} A.,  2008a,
  \mn@doi [\apjl] {10.1086/527355}, \href
  {http://adsabs.harvard.edu/abs/2008ApJ...673L..35A} {673, L35}

\bibitem[\protect\citeauthoryear{{Altamirano}, {van der Klis}, {M{\'e}ndez},
  {Jonker}, {Klein-Wolt}  \& {Lewin}}{{Altamirano}
  et~al.}{2008b}]{Altamirano08a}
{Altamirano} D.,  {van der Klis} M.,  {M{\'e}ndez} M.,  {Jonker} P.~G.,
  {Klein-Wolt} M.,   {Lewin} W.~H.~G.,  2008b, \mn@doi [\apj] {10.1086/590897},
  \href {http://adsabs.harvard.edu/abs/2008ApJ...685..436A} {685, 436}

\bibitem[\protect\citeauthoryear{{Antia} et~al.,}{{Antia}
  et~al.}{2017}]{Antia17}
{Antia} H.~M.,  et~al., 2017, \mn@doi [\apjs] {10.3847/1538-4365/aa7a0e}, \href
  {http://adsabs.harvard.edu/abs/2017ApJS..231...10A} {231, 10}

\bibitem[\protect\citeauthoryear{{Arnaud}}{{Arnaud}}{1996}]{Arnaud96}
{Arnaud} K.~A.,  1996, in {Jacoby} G.~H.,  {Barnes} J.,  eds,  Astronomical
  Society of the Pacific Conference Series Vol. 101, Astronomical Data Analysis
  Software and Systems V. p.~17

\bibitem[\protect\citeauthoryear{{Belloni}, {Homan}, {Motta}, {Ratti}  \&
  {M{\'e}ndez}}{{Belloni} et~al.}{2007}]{Belloni07}
{Belloni} T.,  {Homan} J.,  {Motta} S.,  {Ratti} E.,   {M{\'e}ndez} M.,  2007,
  \mn@doi [\mnras] {10.1111/j.1365-2966.2007.11943.x}, \href
  {http://adsabs.harvard.edu/abs/2007MNRAS.379..247B} {379, 247}

\bibitem[\protect\citeauthoryear{{Bhattacharyya}}{{Bhattacharyya}}{2010}]{Sudip10}
{Bhattacharyya} S.,  2010, \mn@doi [Advances in Space Research]
  {10.1016/j.asr.2010.01.010}, \href
  {http://adsabs.harvard.edu/abs/2010AdSpR..45..949B} {45, 949}

\bibitem[\protect\citeauthoryear{{Bhattacharyya} \&
  {Strohmayer}}{{Bhattacharyya} \& {Strohmayer}}{2006}]{Sudip06b}
{Bhattacharyya} S.,  {Strohmayer} T.~E.,  2006, \mn@doi [\apjl]
  {10.1086/500199}, \href {http://adsabs.harvard.edu/abs/2006ApJ...636L.121B}
  {636, L121}

\bibitem[\protect\citeauthoryear{{Bhattacharyya} et~al.,}{{Bhattacharyya}
  et~al.}{2018}]{Sudip18}
{Bhattacharyya} S.,  et~al., 2018, \mn@doi [\apj] {10.3847/1538-4357/aac495},
  \href {http://adsabs.harvard.edu/abs/2018ApJ...860...88B} {860, 88}

\bibitem[\protect\citeauthoryear{{Boirin}, {Keek}, {M{\'e}ndez}, {Cumming},
  {in't Zand}, {Cottam}, {Paerels}  \& {Lewin}}{{Boirin}
  et~al.}{2007}]{Boirin07}
{Boirin} L.,  {Keek} L.,  {M{\'e}ndez} M.,  {Cumming} A.,  {in't Zand}
  J.~J.~M.,  {Cottam} J.,  {Paerels} F.,   {Lewin} W.~H.~G.,  2007, \mn@doi
  [\aap] {10.1051/0004-6361:20066204}, \href
  {http://adsabs.harvard.edu/abs/2007A%26A...465..559B} {465, 559}

\bibitem[\protect\citeauthoryear{{Chen}, {Zhang}, {Zhang}, {Ji}, {Torres},
  {Kretschmar}, {Li}  \& {Wang}}{{Chen} et~al.}{2013}]{Chen13}
{Chen} Y.-P.,  {Zhang} S.,  {Zhang} S.-N.,  {Ji} L.,  {Torres} D.~F.,
  {Kretschmar} P.,  {Li} J.,   {Wang} J.-M.,  2013, \mn@doi [\apjl]
  {10.1088/2041-8205/777/1/L9}, \href
  {http://adsabs.harvard.edu/abs/2013ApJ...777L...9C} {777, L9}

\bibitem[\protect\citeauthoryear{{Degenaar}, {Koljonen}, {Chakrabarty}, {Kara},
  {Altamirano}, {Miller}  \& {Fabian}}{{Degenaar} et~al.}{2016}]{Degenaar16}
{Degenaar} N.,  {Koljonen} K.~I.~I.,  {Chakrabarty} D.,  {Kara} E.,
  {Altamirano} D.,  {Miller} J.~M.,   {Fabian} A.~C.,  2016, \mn@doi [\mnras]
  {10.1093/mnras/stv2965}, \href
  {http://adsabs.harvard.edu/abs/2016MNRAS.456.4256D} {456, 4256}

\bibitem[\protect\citeauthoryear{{Fisker}, {Thielemann}  \&
  {Wiescher}}{{Fisker} et~al.}{2004}]{Fisker04}
{Fisker} J.~L.,  {Thielemann} F.-K.,   {Wiescher} M.,  2004, \mn@doi [\apjl]
  {10.1086/422215}, \href {http://adsabs.harvard.edu/abs/2004ApJ...608L..61F}
  {608, L61}

\bibitem[\protect\citeauthoryear{{Fisker}, {Schatz}  \& {Thielemann}}{{Fisker}
  et~al.}{2008}]{Fisker08}
{Fisker} J.~L.,  {Schatz} H.,   {Thielemann} F.-K.,  2008, \mn@doi [\apjs]
  {10.1086/521104}, \href {http://adsabs.harvard.edu/abs/2008ApJS..174..261F}
  {174, 261}

\bibitem[\protect\citeauthoryear{{Galloway}, {Muno}, {Hartman}, {Psaltis}  \&
  {Chakrabarty}}{{Galloway} et~al.}{2008}]{Galloway08}
{Galloway} D.~K.,  {Muno} M.~P.,  {Hartman} J.~M.,  {Psaltis} D.,
  {Chakrabarty} D.,  2008, \mn@doi [\apjs] {10.1086/592044}, \href
  {http://adsabs.harvard.edu/abs/2008ApJS..179..360G} {179, 360}

\bibitem[\protect\citeauthoryear{{Grebenev} \& {Chelovekov}}{{Grebenev} \&
  {Chelovekov}}{2017}]{Grebenev17}
{Grebenev} S.~A.,  {Chelovekov} I.~V.,  2017, \mn@doi [Astronomy Letters]
  {10.1134/S106377371709002X}, \href
  {http://adsabs.harvard.edu/abs/2017AstL...43..583G} {43, 583}

\bibitem[\protect\citeauthoryear{{Hasinger} \& {van der Klis}}{{Hasinger} \&
  {van der Klis}}{1989}]{Hasinger89}
{Hasinger} G.,  {van der Klis} M.,  1989, \aap, \href
  {http://adsabs.harvard.edu/abs/1989A%26A...225...79H} {225, 79}

\bibitem[\protect\citeauthoryear{{Heger}, {Cumming}, {Galloway}  \&
  {Woosley}}{{Heger} et~al.}{2007}]{Heger07}
{Heger} A.,  {Cumming} A.,  {Galloway} D.~K.,   {Woosley} S.~E.,  2007, \mn@doi
  [\apjl] {10.1086/525522}, \href
  {http://adsabs.harvard.edu/abs/2007ApJ...671L.141H} {671, L141}

\bibitem[\protect\citeauthoryear{{Kajava}, {S{\'a}nchez-Fern{\'a}ndez},
  {Kuulkers}  \& {Poutanen}}{{Kajava} et~al.}{2017}]{Kajava17}
{Kajava} J.~J.~E.,  {S{\'a}nchez-Fern{\'a}ndez} C.,  {Kuulkers} E.,
  {Poutanen} J.,  2017, \mn@doi [\aap] {10.1051/0004-6361/201629542}, \href
  {http://adsabs.harvard.edu/abs/2017A%26A...599A..89K} {599, A89}

\bibitem[\protect\citeauthoryear{{Keek} \& {Heger}}{{Keek} \&
  {Heger}}{2017}]{Keek17}
{Keek} L.,  {Heger} A.,  2017, \mn@doi [\apj] {10.3847/1538-4357/aa7748}, \href
  {http://adsabs.harvard.edu/abs/2017ApJ...842..113K} {842, 113}

\bibitem[\protect\citeauthoryear{{Keek}, {Galloway}, {in't Zand}  \&
  {Heger}}{{Keek} et~al.}{2010}]{Keek10}
{Keek} L.,  {Galloway} D.~K.,  {in't Zand} J.~J.~M.,   {Heger} A.,  2010,
  \mn@doi [\apj] {10.1088/0004-637X/718/1/292}, \href
  {http://adsabs.harvard.edu/abs/2010ApJ...718..292K} {718, 292}

\bibitem[\protect\citeauthoryear{{Krimm} et~al.,}{{Krimm}
  et~al.}{2013}]{Krimm13}
{Krimm} H.~A.,  et~al., 2013, \mn@doi [\apjs] {10.1088/0067-0049/209/1/14},
  \href {http://adsabs.harvard.edu/abs/2013ApJS..209...14K} {209, 14}

\bibitem[\protect\citeauthoryear{{Kuulkers}, {in't Zand}, {Homan}, {van
  Straaten}, {Altamirano}  \& {van der Klis}}{{Kuulkers}
  et~al.}{2004}]{Kuulkers04}
{Kuulkers} E.,  {in't Zand} J.,  {Homan} J.,  {van Straaten} S.,  {Altamirano}
  D.,   {van der Klis} M.,  2004, in {Kaaret} P.,  {Lamb} F.~K.,   {Swank}
  J.~H.,  eds,  American Institute of Physics Conference Series Vol. 714, X-ray
  Timing 2003: Rossi and Beyond. pp 257--260 (\mn@eprint {}
  {astro-ph/0402076}), \mn@doi{10.1063/1.1781037}

\bibitem[\protect\citeauthoryear{{Lewin}, {van Paradijs}  \& {Taam}}{{Lewin}
  et~al.}{1993}]{Lewin93}
{Lewin} W.~H.~G.,  {van Paradijs} J.,   {Taam} R.~E.,  1993, \mn@doi [\ssr]
  {10.1007/BF00196124}, \href
  {http://adsabs.harvard.edu/abs/1993SSRv...62..223L} {62, 223}

\bibitem[\protect\citeauthoryear{{Linares} et~al.,}{{Linares}
  et~al.}{2009}]{Linares09}
{Linares} M.,  et~al., 2009, The Astronomer's Telegram, \href
  {http://adsabs.harvard.edu/abs/2009ATel.1979....1L} {1979}

\bibitem[\protect\citeauthoryear{{Linares}, {Altamirano}, {Chakrabarty},
  {Cumming}  \& {Keek}}{{Linares} et~al.}{2012}]{Linares12}
{Linares} M.,  {Altamirano} D.,  {Chakrabarty} D.,  {Cumming} A.,   {Keek} L.,
  2012, \mn@doi [\apj] {10.1088/0004-637X/748/2/82}, \href
  {http://adsabs.harvard.edu/abs/2012ApJ...748...82L} {748, 82}

\bibitem[\protect\citeauthoryear{{Maccarone} \& {Coppi}}{{Maccarone} \&
  {Coppi}}{2003}]{Maccarone03}
{Maccarone} T.~J.,  {Coppi} P.~S.,  2003, \mn@doi [\aap]
  {10.1051/0004-6361:20021881}, \href
  {http://adsabs.harvard.edu/abs/2003A%26A...399.1151M} {399, 1151}

\bibitem[\protect\citeauthoryear{{Motta} et~al.,}{{Motta}
  et~al.}{2011}]{Motta11}
{Motta} S.,  et~al., 2011, \mn@doi [\mnras] {10.1111/j.1365-2966.2011.18483.x},
  \href {http://adsabs.harvard.edu/abs/2011MNRAS.414.1508M} {414, 1508}

\bibitem[\protect\citeauthoryear{{Paul}}{{Paul}}{2013}]{Paul13}
{Paul} B.,  2013, \mn@doi [International Journal of Modern Physics D]
  {10.1142/S0218271813410095}, \href
  {http://adsabs.harvard.edu/abs/2013IJMPD..2241009P} {22, 1341009}

\bibitem[\protect\citeauthoryear{{Rao} et~al.,}{{Rao} et~al.}{2017}]{Rao17}
{Rao} A.~R.,  et~al., 2017, \mn@doi [Journal of Astrophysics and Astronomy]
  {10.1007/s12036-017-9450-0}, \href
  {http://adsabs.harvard.edu/abs/2017JApA...38...33R} {38, 33}

\bibitem[\protect\citeauthoryear{{Schatz} et~al.,}{{Schatz}
  et~al.}{1998}]{Schatz98}
{Schatz} H.,  et~al., 1998, \mn@doi [\physrep] {10.1016/S0370-1573(97)00048-3},
  \href {http://adsabs.harvard.edu/abs/1998PhR...294..167S} {294}

\bibitem[\protect\citeauthoryear{{Strohmayer} \& {Bildsten}}{{Strohmayer} \&
  {Bildsten}}{2006}]{Strohmayer06}
{Strohmayer} T.,  {Bildsten} L.,  2006, {New views of thermonuclear bursts}.
pp 113--156

\bibitem[\protect\citeauthoryear{{Strohmayer} \& {Markwardt}}{{Strohmayer} \&
  {Markwardt}}{2002}]{Strohmayer02}
{Strohmayer} T.~E.,  {Markwardt} C.~B.,  2002, \mn@doi [\apj] {10.1086/342152},
  \href {http://adsabs.harvard.edu/abs/2002ApJ...577..337S} {577, 337}

\bibitem[\protect\citeauthoryear{{Sztajno}, {van Paradijs}, {Lewin}, {Trumper},
  {Stollman}, {Pietsch}  \& {van der Klis}}{{Sztajno} et~al.}{1985}]{Sztajno85}
{Sztajno} M.,  {van Paradijs} J.,  {Lewin} W.~H.~G.,  {Trumper} J.,  {Stollman}
  G.,  {Pietsch} W.,   {van der Klis} M.,  1985, \mn@doi [\apj]
  {10.1086/163715}, \href {http://adsabs.harvard.edu/abs/1985ApJ...299..487S}
  {299, 487}

\bibitem[\protect\citeauthoryear{{Verdhan Chauhan} et~al.,}{{Verdhan Chauhan}
  et~al.}{2017}]{Verdhan17}
{Verdhan Chauhan} J.,  et~al., 2017, \mn@doi [\apj] {10.3847/1538-4357/aa6d7e},
  \href {http://adsabs.harvard.edu/abs/2017ApJ...841...41V} {841, 41}

\bibitem[\protect\citeauthoryear{{Wijnands}}{{Wijnands}}{2001}]{Wijnands01}
{Wijnands} R.,  2001, \mn@doi [\apjl] {10.1086/320922}, \href
  {http://adsabs.harvard.edu/abs/2001ApJ...554L..59W} {554, L59}

\bibitem[\protect\citeauthoryear{{Wilms}, {Allen}  \& {McCray}}{{Wilms}
  et~al.}{2000}]{Wilms00}
{Wilms} J.,  {Allen} A.,   {McCray} R.,  2000, \mn@doi [\apj] {10.1086/317016},
  \href {http://adsabs.harvard.edu/abs/2000ApJ...542..914W} {542, 914}

\bibitem[\protect\citeauthoryear{{Woosley} \& {Taam}}{{Woosley} \&
  {Taam}}{1976}]{Woosley76}
{Woosley} S.~E.,  {Taam} R.~E.,  1976, \mn@doi [\nat] {10.1038/263101a0}, \href
  {http://adsabs.harvard.edu/abs/1976Natur.263..101W} {263, 101}

\bibitem[\protect\citeauthoryear{{Woosley} et~al.,}{{Woosley}
  et~al.}{2004}]{Woosley04}
{Woosley} S.~E.,  et~al., 2004, \mn@doi [\apjs] {10.1086/381533}, \href
  {http://adsabs.harvard.edu/abs/2004ApJS..151...75W} {151, 75}

\bibitem[\protect\citeauthoryear{Yadav et~al.,}{Yadav et~al.}{2016}]{Yadav16}
Yadav J.~S.,  et~al., 2016, \mn@doi [Proc. SPIE] {10.1117/12.2231857}, 9905,
  99051D

\bibitem[\protect\citeauthoryear{{Zhang}, {M{\'e}ndez}, {Altamirano}, {Belloni}
   \& {Homan}}{{Zhang} et~al.}{2009}]{Zhang09}
{Zhang} G.,  {M{\'e}ndez} M.,  {Altamirano} D.,  {Belloni} T.~M.,   {Homan} J.,
   2009, \mn@doi [\mnras] {10.1111/j.1365-2966.2009.15148.x}, \href
  {http://adsabs.harvard.edu/abs/2009MNRAS.398..368Z} {398, 368}

\bibitem[\protect\citeauthoryear{{in 't Zand}}{{in 't Zand}}{2011}]{Zand11}
{in 't Zand} J.,  2011, preprint, \href
  {http://adsabs.harvard.edu/abs/2011arXiv1102.3345I} {} (\mn@eprint {arXiv}
  {1102.3345})

\bibitem[\protect\citeauthoryear{{van Paradijs}, {Sztajno}, {Lewin}, {Trumper},
  {Vacca}  \& {van der Klis}}{{van Paradijs} et~al.}{1986}]{Van86}
{van Paradijs} J.,  {Sztajno} M.,  {Lewin} W.~H.~G.,  {Trumper} J.,  {Vacca}
  W.~D.,   {van der Klis} M.,  1986, \mn@doi [\mnras]
  {10.1093/mnras/221.3.617}, \href
  {http://adsabs.harvard.edu/abs/1986MNRAS.221..617V} {221, 617}

\makeatother
\end{thebibliography}
 \bibliographystyle{mnras}

\newpage
\pagebreak
\appendix

\section{Thermonuclear X-ray Bursts as seen in three
proportional counter units of \textsc{LAXPC}}

Here, we show burst profiles as seen in different 
proportional counters of \textsc{LAXPC}.
The light curves were binned with a binsize of 1~second.
The count-rates and profile shapes are consistent
in all the detectors.

\begin{figure*}
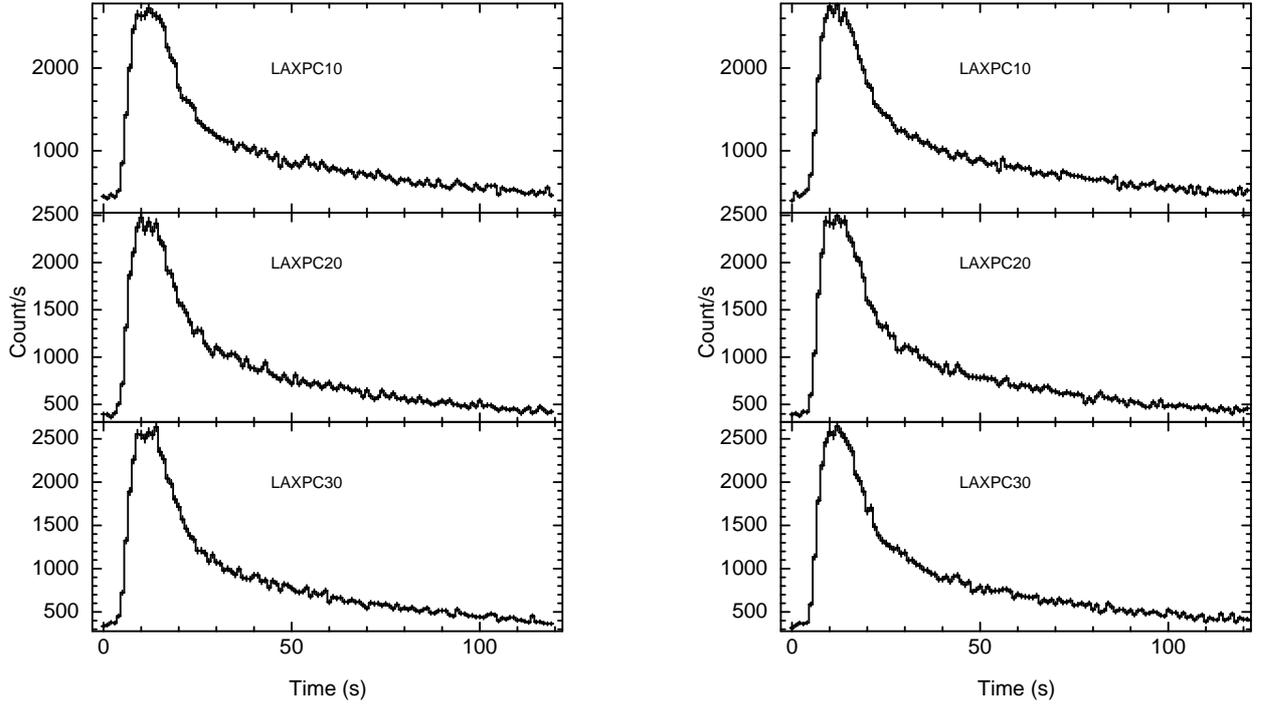

\centering
\begin{minipage}{0.45\textwidth}
\includegraphics[width=\columnwidth]{fig16.ps}
\end{minipage}
\hspace{0.05\linewidth}
\begin{minipage}{0.45\textwidth}
 \includegraphics[width=\columnwidth]{fig17.ps}
 \end{minipage}
\caption{Left:~The first thermonuclear X-ray burst~(burst~1) as seen in the three detectors
of \textsc{LAXPC}.~Right:~The second X-ray burst~(burst~2) observed in this observation of 4U~1636--536.}
\label{Burst1-LC}
\end{figure*}

\begin{figure*}
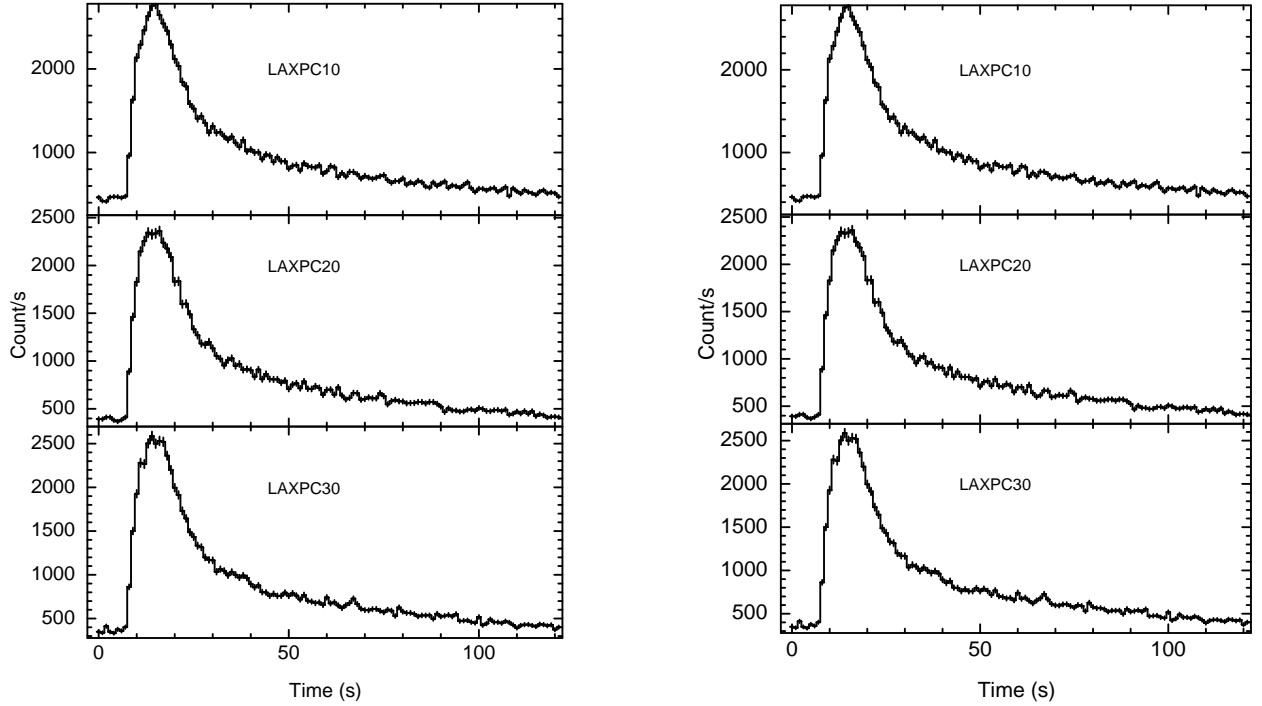

\centering
\begin{minipage}{0.45\textwidth}
\includegraphics[width=\columnwidth]{fig18.ps}
\end{minipage}
\hspace{0.05\linewidth}
\begin{minipage}{0.45\textwidth}
 \includegraphics[width=\columnwidth]{fig19.ps}
\end{minipage}

\caption{Left:~Thermonuclear X-ray burst~(burst~3) as observed with the three detectors of \textsc{LAXPC}.~Right:~
The fourth X-ray burst~(burst~4) as observed with the three detectors of \textsc{LAXPC}.}
\label{Burst2-LC}
\end{figure*}

\begin{figure*}
\centering
\includegraphics[width=\columnwidth]{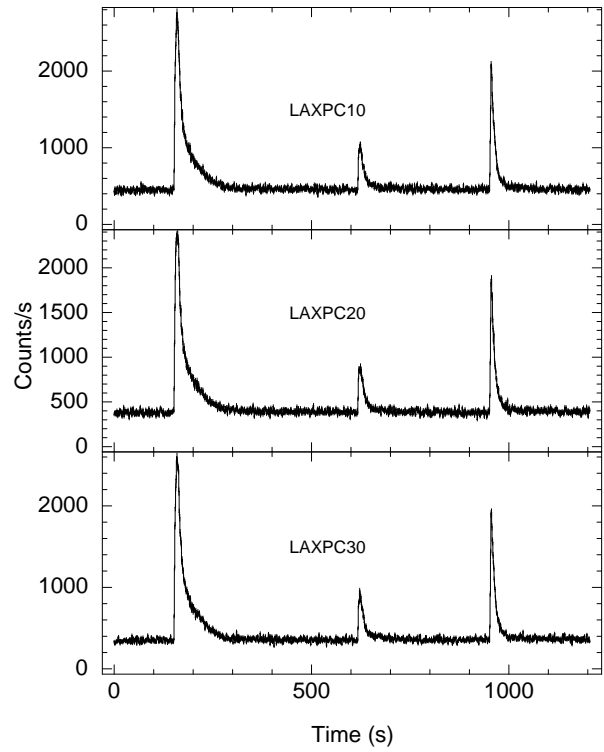}
\caption{The triplet
of X-ray bursts seen in the observation of 4U~1636--536.}
\label{Burst3-LC}
\end{figure*}

\end{document}